\documentclass{PoS}
\usepackage{graphicx,epsfig,color}
\usepackage{wrapfig,rotating}
\usepackage{array}
\usepackage{amsfonts}       
\usepackage{amsmath}        
\usepackage{amssymb}        
\usepackage{hhline}
\usepackage{feynmp}
\usepackage{multirow}
\usepackage{subfigure}
\usepackage{url}
\usepackage{epstopdf}
\AppendGraphicsExtensions{.eps.gz}
\title{SUSY model and dark matter determination in the compressed-spectrum region at the ILC}

\ShortTitle{SUSY models and DM at ILC}

\author{\speaker{Mikael Berggren}\thanks{On behalf of the ILC Physics and Detector Study.}\\
        DESY, 22603 Hamburg, Germany\\
        E-mail: \email{mikael.berggren@dey.de}}

\abstract{
It is an appealing possibility that the observed dark matter density in 
the universe can be fully explained by SUSY. The current experimental 
knowledge indicates that this possibility strongly favors a 
co-annihilation scenario. In such scenarios, the mass difference between 
the next-to-lightest SUSY particle (the NLSP) and the lightest one (the 
LSP) is quite small, which assures that the annihilation cross-section 
is sufficient not to predict a too large abundance of dark matter. 
However, the small mass difference also means that observing SUSY 
becomes hard at hadron colliders, where the observation hinges on the 
tell-tale signature of missing transverse energy: if the mass difference NLSP-to-LSP is 
small, only little energy is carried away by the invisible LSP. This is 
also true even if several other SUSY particles are within the kinematic 
reach, since these states would to a large extent decay via
cascades ending with an NLSP to LSP decay. A lepton collider does not 
have this problem. The clean environment and known initial state at such 
machines assures that SUSY can be detected even if the mass difference 
is very small, provided the center-of-mass energy is sufficiently high. 
We present
prospects for observation and precision characterization of SUSY with 
small mass differences at the ILC, based on detailed simulations
of the ILD detector concept. The resulting possibility to predict the 
dark matter relic density is evaluated and  compared to the 
precision obtained from the Planck mission. Taking a specific model as an 
example, we also discuss the synergies from combining ILC and HL-LHC 
results.
         }

\FullConference{38th International Conference on High Energy Physics\\
		3-10 August 2016\\
		Chicago, USA}

\begin{document}

\def\leqsim{\mathbin{\;\raise1pt\hbox{$<$}\kern-8pt\lower3pt\hbox{$\sim$}\;}}
\def\geqsim{\mathbin{\;\raise1pt\hbox{$>$}\kern-8pt\lower3pt\hbox{$\sim$}\;}}
\renewcommand\topfraction{1.}
\renewcommand\bottomfraction{1.}
\renewcommand\floatpagefraction{0.}
\renewcommand\textfraction{0.}
\def\MXN#1{\mbox{$ M_{\tilde{\chi}^0_#1}                                $}}
\def\MXC#1{\mbox{$ M_{\tilde{\chi}^{\pm}_#1}                            $}}
\def\XP#1{\mbox{$ \tilde{\chi}^+_#1                                     $}}
\def\XM#1{\mbox{$ \tilde{\chi}^-_#1                                     $}}
\def\XPM#1{\mbox{$ \tilde{\chi}^{\pm}_#1                                $}}
\def\XMP#1{\mbox{$ \tilde{\chi}^{\mp}_#1                                $}}
\def\XN#1{\mbox{$ \tilde{\chi}^0_#1                                     $}}
\def\XNN#1#2{\mbox{$ \tilde{\chi}^0_{#1,#2}                             $}}
\def\MXn{\mbox{$ M_{\tilde{\chi}^0}                                     $}}
\def\MXc{\mbox{$ M_{\tilde{\chi}^{\pm}}                                 $}}
\def\Xp{\mbox{$ \tilde{\chi}^+                                          $}}
\def\Xm{\mbox{$ \tilde{\chi}^-                                          $}}
\def\Xpm{\mbox{$ \tilde{\chi}^{\pm}                                     $}}
\def\Xn{\mbox{$ \tilde{\chi}^0                                          $}}
\def\Xnn{\mbox{$ \tilde{\chi}^0                                         $}}
\def\p#1{\mbox{$ \mbox{\bf p}_1                                         $}}
\def\Xgen{\mbox{$ \tilde{\chi}                                          $}}
\def\Mlsp{\mbox{$ M_{\mathrm {LSP}}                                     $}}
\def\lsp{\mbox{$ {\mathrm {LSP}}                                     $}}
\newcommand{\Ptmis}   {\mbox{$/\mkern-11mu P_t \,                          $}}
\newcommand{\Tpmis}   {\mbox{$\theta_{/\mkern-11mu p}                      $}}
\newcommand{\grav}    {\mbox{$ \tilde{\mathrm G}                           $}}
\newcommand{\Gino}    {\mbox{$ \tilde{\mathrm G}                           $}}
\newcommand{\tanb}    {\mbox{$ \tan \beta                                  $}}
\newcommand{\smu}     {\mbox{$ \tilde{\mu}                                 $}}
\newcommand{\smur}    {\mbox{$ \tilde{\mu}_{\mathrm R}                     $}}
\newcommand{\smul}    {\mbox{$ \tilde{\mu}_{\mathrm L}                     $}}
\newcommand{\msmu}    {\mbox{$ M_{\tilde{\mu}}                             $}}
\newcommand{\msmur}   {\mbox{$ M_{\tilde{\mu}_{\mathrm R}}                 $}}
\newcommand{\msmul}   {\mbox{$ M_{\tilde{\mu}_{\mathrm L}}                 $}}
\newcommand{\sel}     {\mbox{$ \tilde{\mathrm e}                           $}}
\newcommand{\sell}    {\mbox{$ \tilde{\mathrm e}_{\mathrm L}               $}}
\newcommand{\selr}    {\mbox{$ \tilde{\mathrm e}_{\mathrm R}               $}}
\newcommand{\msel}    {\mbox{$ M_{\tilde{\mathrm e}}                       $}}
\newcommand{\snu}     {\mbox{$ \tilde\nu                                   $}}
\newcommand{\msnu}    {\mbox{$ m_{\tilde\nu}                               $}}
\newcommand{\msell}   {\mbox{$ M_{\tilde{\mathrm e}_{\mathrm L}}           $}}
\newcommand{\mselr}   {\mbox{$ M_{\tilde{\mathrm e}_{\mathrm R}}           $}}
\newcommand{\fe}      {\mbox{$ \mathrm f                                   $}}
\newcommand{\feb}     {\mbox{$ \overline{\mathrm f}                        $}}
\newcommand{\sfe}     {\mbox{$ \tilde{\mathrm f}                           $}}
\newcommand{\sfeb}    {\mbox{$ \overline{\tilde{\mathrm f}}                $}}
\newcommand{\sfel}    {\mbox{$ \tilde{\mathrm f}_{\mathrm L}               $}}
\newcommand{\sfer}    {\mbox{$ \tilde{\mathrm f}_{\mathrm R}               $}}
\newcommand{\sfelb}   {\mbox{$ \overline{\tilde{\mathrm f}_{\mathrm L}}    $}}
\newcommand{\sferb}   {\mbox{$ \overline{\tilde{\mathrm f}_{\mathrm R}}    $}}
\newcommand{\msfe}    {\mbox{$ M_{\tilde{\mathrm f}}                       $}}
\newcommand{\sle}     {\mbox{$ \tilde{\ell}                                $}}
\newcommand{\sq}     {\mbox{$ \tilde{q}                                $}}
\newcommand{\sqr}     {\mbox{$ \tilde{q}_{\mathrm R}                                $}}
\newcommand{\sql}     {\mbox{$ \tilde{q}_{\mathrm L}                                $}}
\newcommand{\msle}    {\mbox{$ M_{\tilde{\ell}}                            $}}
\newcommand{\stau}    {\mbox{$ \tilde{\tau}                                $}}
\newcommand{\stone}   {\mbox{$ \tilde{\tau}_1                              $}}
\newcommand{\sttwo}   {\mbox{$ \tilde{\tau}_2                              $}}
\newcommand{\staur}   {\mbox{$ \tilde{\tau}_{\mathrm R}                    $}}
\newcommand{\mstau}   {\mbox{$ M_{\tilde{\tau}}                            $}}
\newcommand{\mstone}  {\mbox{$ M_{\tilde{\tau}_1}                          $}}
\newcommand{\msttwo}  {\mbox{$ M_{\tilde{\tau}_2}                          $}}
\newcommand{\stq}     {\mbox{$ \tilde {\mathrm t}                          $}}
\newcommand{\stqone}  {\mbox{$ \tilde {\mathrm t}_1                        $}}
\newcommand{\stqtwo}  {\mbox{$ \tilde {\mathrm t}_2                        $}}
\newcommand{\msq}    {\mbox{$ M_{\tilde {\mathrm q}}                      $}}
\newcommand{\mstq}    {\mbox{$ M_{\tilde {\mathrm t}}                      $}}
\newcommand{\sbq}     {\mbox{$ \tilde {\mathrm b}                          $}}
\newcommand{\sbqone}  {\mbox{$ \tilde {\mathrm b}_1                        $}}
\newcommand{\sbqtwo}  {\mbox{$ \tilde {\mathrm b}_2                        $}}
\newcommand{\msbq}    {\mbox{$ M_{\tilde {\mathrm b}}                      $}}
\newcommand{\msbqone}    {\mbox{$ M_{\tilde {\mathrm b}_1}                 $}}
\newcommand{\msbqtwo}    {\mbox{$ M_{\tilde {\mathrm b}_2}                 $}}
\newcommand{\mstqone}    {\mbox{$ M_{\tilde {\mathrm t}_1}                 $}}
\newcommand{\mstqtwo}    {\mbox{$ M_{\tilde {\mathrm t}_2}                 $}}
\newcommand{\An}      {\mbox{$ {\, \mathrm A}^0                               $}}
\newcommand{\hn}      {\mbox{$ {\, \mathrm h}^0                               $}}
\newcommand{\Zn}      {\mbox{$ {\, \mathrm Z}                                 $}}
\newcommand{\Zstar}   {\mbox{$ {\, \mathrm Z}^*                               $}}
\newcommand{\Hn}      {\mbox{$ {\, \mathrm H}^0                               $}}
\newcommand{\HP}      {\mbox{$ {\, \mathrm H}^+                               $}}
\newcommand{\HM}      {\mbox{$ {\, \mathrm H}^-                               $}}
\newcommand{\W}      {\mbox{$ {\, \mathrm W}                               $}}
\newcommand{\Wp}      {\mbox{$ {\, \mathrm W}^+                               $}}
\newcommand{\Wm}      {\mbox{$ {\, \mathrm W}^-                               $}}
\newcommand{\Wstar}   {\mbox{$ {\, \mathrm W}^*                               $}}
\newcommand{\WW}      {\mbox{$ {\, \mathrm W}^+{\mathrm W}^-                  $}}
\newcommand{\ZZ}      {\mbox{$ {\, \mathrm{Z Z}}$}}
\newcommand{\HZ}      {\mbox{$ {\, \mathrm H}^0 {\mathrm Z}                   $}}
\newcommand{\GW}      {\mbox{$ \Gamma_{\mathrm W}                          $}}
\newcommand{\Zg}      {\mbox{$ \Zn \gamma                                  $}}
\newcommand{\Zorg}      {\mbox{$ \Zn / \gamma                                  $}}
\newcommand{\sqs}     {\mbox{$ \sqrt{s}                                    $}}
\newcommand{\epm}     {\mbox{$ {\, \mathrm e}^{\pm}                           $}}
\newcommand{\ee}      {\mbox{$ {\, \mathrm e}^+ {\mathrm e}^-                 $}}
\newcommand{\mumu}    {\mbox{$ \, \mu^+ \mu^-                                 $}}
\newcommand{\tautau}  {\mbox{$ \, \tau^+ \tau^-                               $}}
\newcommand{\eeto}    {\mbox{$ {\, \mathrm e}^+ {\mathrm e}^- \to             $}}
\newcommand{\ellell}  {\mbox{$ \, \ell^+ \ell^-                               $}}
\newcommand{\eeWW}    {\mbox{$ \, \ee \rightarrow \, \WW                         $}}
\newcommand{\MeV}     {\mbox{$ {\mathrm{MeV}}                              $}}
\newcommand{\MeVc}    {\mbox{$ {\mathrm{MeV}}/c                            $}}
\newcommand{\MeVcc}   {\mbox{$ {\mathrm{MeV}}/c^2                          $}}
\newcommand{\GeV}     {\mbox{$ {\mathrm{GeV}}                              $}}
\newcommand{\GeVc}    {\mbox{$ {\mathrm{GeV}}/c                            $}}
\newcommand{\GeVcc}   {\mbox{$ {\mathrm{GeV}}/c^2                          $}}
\newcommand{\TeV}     {\mbox{$ {\mathrm{TeV}}                              $}}
\newcommand{\TeVc}    {\mbox{$ {\mathrm{TeV}}/c                            $}}
\newcommand{\TeVcc}   {\mbox{$ {\mathrm{TeV}}/c^2                          $}}
\newcommand{\pbi}     {\mbox{$ {\mathrm{pb}}^{-1}                          $}}
\newcommand{\MZ}      {\mbox{$ m_{{\mathrm Z}}                             $}}
\newcommand{\MW}      {\mbox{$ m_{\mathrm W}                               $}}
\newcommand{\MA}      {\mbox{$ m_{\mathrm A}                               $}}
\newcommand{\GF}      {\mbox{$ {\mathrm G}_{\mathrm F}                     $}}
\newcommand{\MH}      {\mbox{$ m_{{\mathrm H}^0}                           $}}
\newcommand{\MHP}     {\mbox{$ m_{{\mathrm H}^\pm}                         $}}
\newcommand{\MSH}     {\mbox{$ m_{{\mathrm h}^0}                           $}}
\newcommand{\MT}      {\mbox{$ m_{\mathrm t}                               $}}
\newcommand{\GZ}      {\mbox{$ \Gamma_{{\mathrm Z} }                       $}}

\newcommand{\TT}      {\mbox{$ \mathrm T                                   $}}
\newcommand{\UU}      {\mbox{$ \mathrm U                                   $}}
\newcommand{\alphmz}  {\mbox{$ \alpha (m_{{\mathrm Z}})                    $}}
\newcommand{\alphas}  {\mbox{$ \alpha_{\mathrm s}                          $}}
\newcommand{\alphmsb} {\mbox{$ \alphas (m_{\mathrm Z})
                               _{\overline{\mathrm{MS}}}                   $}}
\newcommand{\alphbar} {\mbox{$ \overline{\alpha}_{\mathrm s}               $}}
\newcommand{\Ptau}    {\mbox{$ P_{\tau}                                    $}}
\newcommand{\mean}[1] {\mbox{$ \left\langle #1 \right\rangle               $}}
\newcommand{\dgree}   {\mbox{$ ^\circ                                      $}}
\newcommand{\qqg}     {\mbox{$ {\mathrm q}\bar{\mathrm q}\gamma            $}}
\newcommand{\Wev}     {\mbox{$ W e \, \nu_e              $}}
\newcommand{\Zvv}     {\mbox{$ \Zn \nu \bar{\nu}                           $}}
\newcommand{\Zee}     {\mbox{$ Z^0 e^+ e^-                                 $}}
\newcommand{\ctw}     {\mbox{$ \cos\theta_{\mathrm W}                      $}}
\newcommand{\thw}     {\mbox{$ \theta_{\mathrm W}                          $}}
\newcommand{\thetabar}{\mbox{$ \theta^*                                    $}}
\newcommand{\phibar}  {\mbox{$ \phi^*                                      $}}
\newcommand{\thetapl} {\mbox{$ \theta_+                                    $}}
\newcommand{\phipl}   {\mbox{$ \phi_+                                      $}}
\newcommand{\thetamin}{\mbox{$ \theta_-                                    $}}
\newcommand{\phimin}  {\mbox{$ \phi_-                                      $}}
\newcommand{\ds}      {\mbox{$ {\mathrm d} \sigma                          $}}
\def    \ll           {\mbox{$\ell \ell                                    $}}
\def    \jjl          {\mbox{$j j \ell                           $}}
\def    \jj           {\mbox{$\jmath \jmath                                $}}
\def   \jjjj          {\mbox{${\it jets}                                   $}}
\newcommand{\jjlv}    {\mbox{$ j j \ell \nu                                $}}
\newcommand{\jjvv}    {\mbox{$ j j \nu \bar{\nu}                           $}}
\newcommand{\qqvv}    {\mbox{$ \mathrm{q \bar{q}} \nu \bar{\nu}            $}}
\newcommand{\qqll}    {\mbox{$ \mathrm{q \bar{q}} \ell \bar{\ell}          $}}
\newcommand{\jjll}    {\mbox{$ j j \ell \bar{\ell}                         $}}
\newcommand{\lvlv}    {\mbox{$ \ell \nu \ell \nu                           $}}
\newcommand{\dz}      {\mbox{$ \delta g_{\mathrm{W W Z}    }               $}}
\newcommand{\pT}      {\mbox{$ p_{\mathrm{T}}                              $}}
\newcommand{\pt}      {\mbox{$ p_{\mathrm{t}}                              $}}
\newcommand{\ptr}     {\mbox{$ p_{\perp}                                   $}}
\newcommand{\ptrjet}  {\mbox{$ p_{\perp {\mathrm{jet}}}                    $}}
\newcommand{\Wvis}    {\mbox{$ {\mathrm W}_{\mathrm{vis}}                  $}}
\newcommand{\gamgam}  {\mbox{$ \gamma \gamma                               $}}
\newcommand{\qaqb}    {\mbox{$ {\, \mathrm q}_1 \,  \bar{\mathrm q}_2      $}}
\newcommand{\qcqd}    {\mbox{$ {\, \mathrm q}_3  \, \bar{\mathrm q}_4      $}}
\newcommand{\bbbar}   {\mbox{$ {\, \mathrm b} \, \bar{\mathrm b}           $}}
\newcommand{\ccbar}   {\mbox{$ {\, \mathrm c} \, \bar{\mathrm c}           $}}
\newcommand{\ffbar}   {\mbox{$ {\, \mathrm f} \, \bar{\mathrm f}           $}}
\newcommand{\ffbarp}  {\mbox{$ {\, \mathrm f} \, \bar{\mathrm f}'          $}}
\newcommand{\qqbar}   {\mbox{$\mathrm q \, \bar{\mathrm q}                 $}}
\newcommand{\quark}   {\mbox{$\mathrm q                                    $}}
\newcommand{\charm}   {\mbox{$\mathrm c                                    $}}
\newcommand{\bottom}  {\mbox{$\mathrm b                                    $}}
\newcommand{\topq}    {\mbox{$\mathrm t                                    $}}
\newcommand{\quarkb}   {\mbox{$\bar{\mathrm q}               $}}
\newcommand{\charmb}   {\mbox{$\bar{\mathrm c}               $}}
\newcommand{\bottomb}  {\mbox{$\bar{\mathrm b}               $}}
\newcommand{\topqb}    {\mbox{$\bar{\mathrm t}               $}}
\newcommand{\nunubar} {\mbox{$ {\, \nu} \, \bar{\nu}                       $}}
\newcommand{\qqbarp}  {\mbox{$ {\, \mathrm q'} \, \bar{\mathrm q}'         $}}
\newcommand{\djoin}   {\mbox{$ d_{\mathrm{join}}                           $}}
\newcommand{\mErad}   {\mbox{$ \left\langle E_{\mathrm{rad}} \right\rangle $}}
\newcommand{\bptre}{\rm b^{+}_{3}}
\newcommand{\bp}{\rm b^{+}_{1}}
\newcommand{\bo}{\rm b^0}
\newcommand{\bos}{\rm b^0_s}
\newcommand{\bss}{\rm b^s_s}
\newcommand{\BsDmX}{{B_{s}^{0}} \rightarrow D \mu X}
\newcommand{\BsDsm}{{B_{s}^{0}} \rightarrow D_{s} \mu X}
\newcommand{\BsDsX}{{B_{s}^{0}} \rightarrow D_{s} X}
\newcommand{\BDsX}{B \rightarrow D_{s} X}
\newcommand{\BDomX}{B \rightarrow D^{0} \mu X}
\newcommand{\BDpmX}{B \rightarrow D^{+} \mu X}
\newcommand{\Dsfmn}{D_{s} \rightarrow \phi \mu \nu}
\newcommand{\Dsfipi}{D_{s} \rightarrow \phi \pi}
\newcommand{\DsfX}{D_{s} \rightarrow \phi X}
\newcommand{\DpfX}{D^{+} \rightarrow \phi X}
\newcommand{\DofX}{D^{0} \rightarrow \phi X}
\newcommand{\DfX}{D \rightarrow \phi X}
\newcommand{\DsD}{B \rightarrow D_{s} D}
\newcommand{\DsmX}{D_{s} \rightarrow \mu X}
\newcommand{\DmX}{D \rightarrow \mu X}
\newcommand{\Zbb}{Z^{0} \rightarrow \rm b \overline{b}}
\newcommand{\Zcc}{Z^{0} \rightarrow \rm c \overline{c}}
\newcommand{\Rbb}{\frac{\Gamma_{Z^0 \rightarrow \rm b \overline{b}}}
{\Gamma_{Z^0 \rightarrow Hadrons}}}
\newcommand{\Rcc}{\frac{\Gamma_{Z^0 \rightarrow \rm c \overline{c}}}
{\Gamma_{Z^0 \rightarrow Hadrons}}}
\newcommand{\str}{\rm s \overline{s}}
\newcommand{\Bs}{\rm{B^0_s}}
\newcommand{\Bsb}{\overline{\rm{B^0_s}}}
\newcommand{\Bp}{\rm{B^{+}}}
\newcommand{\Bm}{\rm{B^{-}}}
\newcommand{\Bo}{\rm{B^{0}}}
\newcommand{\Bd}{\rm{B^{0}_{d}}}
\newcommand{\Bdb}{\overline{\rm{B^{0}_{d}}}}
\newcommand{\Lb}{\Lambda^0_b}
\newcommand{\Lbb}{\overline{\Lambda^0_b}}
\newcommand{\Kstar}{\rm{K^{\star 0}}}
\newcommand{\phim}{\rm{\phi}}
\newcommand{\Ds}{\mbox{D}_s}
\newcommand{\Dsp}{\mbox{D}_s^+}
\newcommand{\Dp}{\mbox{D}^+}
\newcommand{\Dn}{\mbox{D}^0}
\newcommand{\Dsb}{\overline{\mbox{D}_s}}
\newcommand{\Dm}{\mbox{D}^-}
\newcommand{\Dnb}{\overline{\mbox{D}^0}}
\newcommand{\Lc}{\Lambda_c}
\newcommand{\Lcb}{\overline{\Lambda_c}}
\newcommand{\Dstarp}{\mbox{D}^{\ast +}}
\newcommand{\Dstarm}{\mbox{D}^{\ast -}}
\newcommand{\Dsstarp}{\mbox{D}_s^{\ast +}}
\newcommand{\Pb}{P_{b-baryon}}
\newcommand{\KKpi}{\rm{ K K \pi }}
\newcommand{\nb}{\rm{nb}}
\newcommand{\Gm}{\rm{G_{\mu}}}
\newcommand{\Afb}{\rm{A_{FB}}}
\newcommand{\Afbs}{\rm{A_{FB}^{s}}}
\newcommand{\sigmaf}{\sigma_{\rm{F}}}
\newcommand{\sigmab}{\sigma_{\rm{B}}}
\newcommand{\NF}{\rm{N_{F}}}
\newcommand{\NB}{\rm{N_{B}}}
\newcommand{\Nnu}{\rm{N_{\nu}}}
\newcommand{\RZ}{\rm{R_Z}}
\newcommand{\rhob}{\rho_{eff}}
\newcommand{\Gammanz}{\rm{\Gamma_{Z}^{new}}}
\newcommand{\Gammani}{\rm{\Gamma_{inv}^{new}}}
\newcommand{\Gammasz}{\rm{\Gamma_{Z}^{SM}}}
\newcommand{\Gammasi}{\rm{\Gamma_{inv}^{SM}}}
\newcommand{\Gammaxz}{\rm{\Gamma_{Z}^{exp}}}
\newcommand{\Gammaxi}{\rm{\Gamma_{inv}^{exp}}}
\newcommand{\rhoZ}{\rho_{\rm Z}}
\newcommand{\swsq}{\sin^2\!\thw}
\newcommand{\swsqmsb}{\sin^2\!\theta_{\rm W}^{\overline{\rm MS}}}
\newcommand{\swsqbar}{\sin^2\!\overline{\theta}_{\rm W}}
\newcommand{\cwsqbar}{\cos^2\!\overline{\theta}_{\rm W}}
\newcommand{\swsqb}{\sin^2\!\theta^{eff}_{\rm W}}
\newcommand{\eeX}{{e^+e^-X}}
\newcommand{\gaga}{{\gamma\gamma}}
\newcommand{\eeg}{{e^+e^-\gamma}}
\newcommand{\mumug}{{\mu^+\mu^-\gamma}}
\newcommand{\qqb}{{q\bar{q}}}
\newcommand{\eegg}{e^+e^-\rightarrow \gamma\gamma}
\newcommand{\eeggg}{e^+e^-\rightarrow \gamma\gamma\gamma}
\newcommand{\eeee}{e^+e^-\rightarrow e^+e^-}
\newcommand{\eeeeee}{e^+e^-\rightarrow e^+e^-e^+e^-}
\newcommand{\eeeeg}{e^+e^-\rightarrow e^+e^-(\gamma)}
\newcommand{\eeeegg}{e^+e^-\rightarrow e^+e^-\gamma\gamma}
\newcommand{\eeeg}{e^+e^-\rightarrow (e^+)e^-\gamma}
\newcommand{\eemumu}{e^+e^-\rightarrow \mu^+\mu^-}
\newcommand{\eetautau}{e^+e^-\rightarrow \tau^+\tau^-}
\newcommand{\eehad}{e^+e^-\rightarrow {\rm hadrons}}
\newcommand{\eettg}{e^+e^-\rightarrow \tau^+\tau^-\gamma}
\newcommand{\eell}{e^+e^-\rightarrow l^+l^-}
\newcommand{\Ztopig}{{\rm Z}^0\rightarrow \pi^0\gamma}
\newcommand{\Ztogg}{{\rm Z}^0\rightarrow \gamma\gamma}
\newcommand{\Ztoee}{{\rm Z}^0\rightarrow e^+e^-}
\newcommand{\Ztoggg}{{\rm Z}^0\rightarrow \gamma\gamma\gamma}
\newcommand{\Ztomumu}{{\rm Z}^0\rightarrow \mu^+\mu^-}
\newcommand{\Ztotautau}{{\rm Z}^0\rightarrow \tau^+\tau^-}
\newcommand{\Ztoll}{{\rm Z}^0\rightarrow l^+l^-}
\newcommand{\Ztocc}{{\rm Z^0\rightarrow c \bar c}}
\newcommand{\Lamp}{\Lambda_{+}}
\newcommand{\Lamm}{\Lambda_{-}}
\newcommand{\Pt}{\rm P_{t}}
\newcommand{\Gee}{\Gamma_{ee}}
\newcommand{\Gpig}{\Gamma_{\pi^0\gamma}}
\newcommand{\Ggg}{\Gamma_{\gamma\gamma}}
\newcommand{\Gggg}{\Gamma_{\gamma\gamma\gamma}}
\newcommand{\Gmumu}{\Gamma_{\mu\mu}}
\newcommand{\Gtautau}{\Gamma_{\tau\tau}}
\newcommand{\Ginv}{\Gamma_{\rm inv}}
\newcommand{\Ghad}{\Gamma_{\rm had}}
\newcommand{\Gnu}{\Gamma_{\nu}}
\newcommand{\GnuSM}{\Gamma_{\nu}^{\rm SM}}
\newcommand{\Gll}{\Gamma_{l^+l^-}}
\newcommand{\Gff}{\Gamma_{f\overline{f}}}
\newcommand{\Gtot}{\Gamma_{\rm tot}}
\newcommand{\Rb}{\mbox{R}_b}
\newcommand{\Rc}{\mbox{R}_c}
\newcommand{\al}{a_l}
\newcommand{\vl}{v_l}
\newcommand{\af}{a_f}
\newcommand{\vf}{v_f}
\newcommand{\ael}{a_e}
\newcommand{\ve}{v_e}
\newcommand{\amu}{a_\mu}
\newcommand{\vmu}{v_\mu}
\newcommand{\atau}{a_\tau}
\newcommand{\vtau}{v_\tau}
\newcommand{\ahatl}{\hat{a}_l}
\newcommand{\vhatl}{\hat{v}_l}
\newcommand{\ahate}{\hat{a}_e}
\newcommand{\vhate}{\hat{v}_e}
\newcommand{\ahatmu}{\hat{a}_\mu}
\newcommand{\vhatmu}{\hat{v}_\mu}
\newcommand{\ahattau}{\hat{a}_\tau}
\newcommand{\vhattau}{\hat{v}_\tau}
\newcommand{\vtildel}{\tilde{\rm v}_l}
\newcommand{\avsq}{\ahatl^2\vhatl^2}
\newcommand{\Ahatl}{\hat{A}_l}
\newcommand{\Vhatl}{\hat{V}_l}
\newcommand{\Afer}{A_f}
\newcommand{\Ael}{A_e}
\newcommand{\Aferb}{\bar{A_f}}
\newcommand{\Aelb}{\bar{A_e}}
\newcommand{\AVsq}{\Ahatl^2\Vhatl^2}
\newcommand{\Iwk}{I_{3l}}
\newcommand{\Qch}{|Q_{l}|}
\newcommand{\roots}{\sqrt{s}}
\newcommand{\mt}{m_t}
\newcommand{\Rechi}{{\rm Re} \left\{ \chi (s) \right\}}
\newcommand{\up}{^}
\newcommand{\abscosthe}{|cos\theta|}
\newcommand{\dsum}{\Sigma |d_\circ|}
\newcommand{\zsum}{\Sigma z_\circ}
\newcommand{\sint}{\mbox{$\sin\theta$}}
\newcommand{\cost}{\mbox{$\cos\theta$}}
\newcommand{\mcost}{|\cos\theta|}
\newcommand{\epair}{\mbox{$e^{+}e^{-}$}}
\newcommand{\mupair}{\mbox{$\mu^{+}\mu^{-}$}}
\newcommand{\taupair}{\mbox{$\tau^{+}\tau^{-}$}}
\newcommand{\fullskip}{\vskip 16cm}
\newcommand{\halfskip}{\vskip  8cm}
\newcommand{\quarskip}{\vskip  6cm}
\newcommand{\abitskip}{\vskip 0.5cm}
\newcommand{\ba}{\begin{array}}
\newcommand{\ea}{\end{array}}
\newcommand{\bc}{\begin{center}}
\newcommand{\ec}{\end{center}}
\newcommand{\be}{\begin{eqnarray}}
\newcommand{\eeq}{\end{eqnarray}}
\newcommand{\bes}{\begin{eqnarray*}}
\newcommand{\ees}{\end{eqnarray*}}
\newcommand{\Kz}{\ifmmode {\rm K^0_s} \else ${\rm K^0_s} $ \fi}
\newcommand{\Zz}{\ifmmode {\rm Z^0} \else ${\rm Z^0 } $ \fi}
\newcommand{\xxbar}{\ifmmode {\rm x\bar{x}} \else ${\rm x\bar{x}} $ \fi}
\newcommand{\rphi}{\ifmmode {\rm R\phi} \else ${\rm R\phi} $ \fi}

\newcommand{\Lum}{${\cal L}\;$}
\newcommand{\lum}{{\cal L}}
\newcommand{\Cms}{$\mbox{ cm}^{-2} \mbox{ s}^{-1}\;$}
\newcommand{\cms}{\mbox{ cm}^{-2} \mbox{ s}^{-1}\;}
\newcommand{\Ecms}    {\mbox{$ E_{\mathrm{\small cms}}                      $}}
\newcommand{\Ecm}    {\mbox{$ E_{\mathrm{\small cm}}                      $}}
\newcommand{\Etvis}    {\mbox{$ E^{T}_{\mathrm{\small vis}}                      $}}
\newcommand{\Evis}    {\mbox{$ E_{\mathrm{\small vis}}                      $}}
\newcommand{\Erad}    {\mbox{$ E_{\mathrm{\small rad}}                      $}}
\newcommand{\Mvis}    {\mbox{$ M_{\mathrm{\small vis}}                      $}}
\newcommand{\pvis}    {\mbox{$ p_{\mathrm{\small vis}}                      $}}
\newcommand{\Minv}    {\mbox{$ M_{\mathrm{\small inv}}                      $}}
\newcommand{\pmiss}   {\mbox{$ p_{\mathrm{\small miss}}                     $}}
\newcommand{\Mhfit}{\; \hat{m}_{H^0} }
\newcommand{\bl}      {\mbox{\ \ \ \ \ \ \ \ \ \ } }
\newcommand{\Zto}   {\mbox{$\mathrm Z^0 \to$}}
\newcommand{\etal}  {\mbox{\it et al.}}
\def\NPB#1#2#3{{\rm Nucl.~Phys.} {\bf{B#1}} (#2) #3}
\def\PLB#1#2#3{{\rm Phys.~Lett.} {\bf{B#1}} (#2) #3}
\def\PRD#1#2#3{{\rm Phys.~Rev.} {\bf{D#1}} (#2) #3}
\def\PRL#1#2#3{{\rm Phys.~Rev.~Lett.} {\bf{#1}} (#2) #3}
\def\ZPC#1#2#3{{\rm Z.~Phys.} {\bf C#1} (#2) #3}
\def\PTP#1#2#3{{\rm Prog.~Theor.~Phys.} {\bf#1}  (#2) #3}
\def\MPL#1#2#3{{\rm Mod.~Phys.~Lett.} {\bf#1} (#2) #3}
\def\PR#1#2#3{{\rm Phys.~Rep.} {\bf#1} (#2) #3}
\def\RMP#1#2#3{{\rm Rev.~Mod.~Phys.} {\bf#1} (#2) #3}
\def\HPA#1#2#3{{\rm Helv.~Phys.~Acta} {\bf#1} (#2) #3}
\def\NIMA#1#2#3{{\rm Nucl.~Instr.~and~Meth.} {\bf#1} (#2) #3} 
\def\CPC#1#2#3{{\rm Comp.~Phys.~Comm.} {\bf#1} (#2) #3}
\def\EPJC#1#2#3{{\rm E.~Phys.~J.} {\bf{C#1}} (#2) #3}
\def    \DM          {\mbox{$\Delta$M}}
\def    \missEt      {\ifmmode{/\mkern-11mu E_t}\else{${/\mkern-11mu E_t}$}\fi}
\def    \missE       {\ifmmode{/\mkern-11mu E}\else{${/\mkern-11mu E}$}\fi}
\def    \missp       {\ifmmode{/\mkern-11mu p}\else{${/\mkern-11mu p}$}\fi}
\def    \misspt      {\ifmmode{/\mkern-11mu p_t}\else{${/\mkern-11mu p_t}$}\fi}
\def    \DML         {\mbox{5~GeV $<\Delta M<$ 10~GeV}}
\def    \rs          {\mbox{$\sqrt{s}$}}
\def    \msneu       {\mbox{$m_{\tilde{\nu}}$}}
\def    \rad         {\mbox{$\it{rad}$}}

\section{Introduction}
Many observations indicate that a large fraction of the matter in the universe is
dark, i.e. not visible in observations of electromagnetic radiation.
The data on the temperature power spectrum of the 
cosmic microwave background (CMB) and its polarisation 
obtained by the Planck mission give the most precise cosmological determination of the dark matter (DM) relic density $\Omega_{CDM}$, 
and is
$\Omega_{CDM}=0.1197 \pm 0.0022$ \cite{Ade:2015xua}.
If such an abundance of DM in the early universe is to be explained
by the existence of hitherto unknown elementary particle,
one needs a balance between early universe production and
decay of this particle.
One finds that this could be obtained by a weakly interacting, stable
massive particle (a WIMP).
Since the lightest Super-symmetric particle (the LSP) has such properties,
it is quite compelling to assume that it is the DM.
However, if the LSP is the only SUSY particle in the mass-range needed,
it would tend to be too abundant.
Therefore, other SUSY particles are expected to have been close in density
in the hot early universe, i.e. to be close to mass-degenerate with the LSP. 
Such scenarios are called a coannihilation scenarios, 
and the  coannihilation next-to-lightest SUSY particle (the NLSP) could be the $\stone$, the $\stqone$ or the $\XPM{1}$.
Such scenarios are also compelling for other reasons.
If the totality of current observations sensitive to SUSY are
fitted to a general SUSY model with 10 free parameters, as has been
done by the {\tt Mastercode} collaboration\cite{deVries:2015hva}, one finds that
models where the bosino and slepton-sector is quite compressed fits the
observations best.

In this contribution, we study in particular the $\stau$ coannihilation scenario
at the ILC.
Using the {\tt micrOMEGAs} code\cite{Belanger:2006is}, it has been shown \cite{Lehtinen:2016qis}
that a 1\% variation of $\mstau$ or $\MXN{1}$ changes
the predicted relic     abundance by 5 \%, while
a similar variation of the $\stau$ mixing angle ($\theta_{\stau}$) or the binoness of the LSP ($N_{11}$)
changes the abundance by 1\% and 3.5 \%, respectively.
Therefore,  
to match the 2~\% uncertainty on the
relic abundance obtained by Planck
one needs 
per mil-level  precision measurements of the LSP and NLSP masses, 
and percent-level precision on the LSP and NLSP mixings.
In Section 2, we introduce the ILC and point out important differences to the LHC
for the study of the $\stau$ coannihilation scenario.
Section 3 presents the results from our study  on the prospects to measure the key-parameters of the model
at the ILC, and comment on the LHC prospects.


\section{Compressed spectra at the ILC and the LHC}
\label{ILCandLHC}
The International Linear Collider (ILC) \cite{Adolphsen:2013kya} is a proposed 34 km long  $e^+e^-$ collider,
colliding electrons and positrons accelerated by two  separate superconducting linear
accelerators.
The centre-of-mass energy ($E_{CMS}$) is to be tunable between 200 and 500 GeV, with
an upgrade path to 1 TeV defined. 
The electron and positron beams are polarised to 80\% and $>$30\%, respectively\footnote{
In the following, $\mathcal{P}_{p^-,p^+}$ denotes the beam-polarisation configuration
$\mathcal{P}(e^-,e^+)=(p^-,p^+)$, with $p^-$ and $p^+$ given in percent.}.
Once it is tuned up, 
the machine will deliver an integrated luminosity 
of  290 fb$^{-1}$ per year at $E_{CMS}=$ 500 \GeV.

In contrast to proton colliders, it is a collider of fundamental particles, 
meaning that the initial state is known,
and that there is no spectator event accompanying the hard interaction.
Furthermore, the production processes are electroweak, which implies 
lower 
event-rates.
The detectors can be made much thinner, since they are not required to be radiation-hard.
They can be as close as 1 to 2 cm to the interaction-point, and can have very close
to 4$\pi$ coverage.
In addition, the low rates means that the detectors can register all interactions,
i.e. there is no need for any triggering.
In the coannihilation scenario, both the  excellent hermeticity and the trigger-less operation are
required features be able to detect NLSP-pair production events above background from
$\gamma\gamma$ events.
Further details of the proposed detectors for the ILC can be found in \cite{Behnke:2013lya}.

LHC puts strong limits on the mass of first and second generation squarks  and on the gluino.
These states have no direct influence on DM, the muon magnetic moment anomaly, nor
naturalness.
If one requires GUT-scale unification of the SUSY mass-parameters,
the limits do indirectly constrain the states that are relevant, i.e. the third generation squarks 
and the EW sector.
By removing this requirement
\footnote{
The {\it couplings} at the GUT scale can still be required to
unify.},
the LHC limits no longer constrain 
SUSY contributions to the precision observables and DM.
If, in addition, the SUSY spectrum is compressed, one would expect  long decay-cascades at the LHC, ending up at
the NLSP, which 
decays into the LSP and, due to the low mass difference required
in coannihilation models, to  a soft visible system.
This is thus not  a  model predicting large
missing E$_T$, nor a simplified model where the produced heavy coloured particles decay directly
to the LSP.
Taken together, the current limits on  coannihilation models from LHC become
significantly weaker than those obtained from simplified models or large
missing E$_T$ ones.
\section{The Stau-coannihilation STCx models }

\begin{figure}[htb]
  \begin{center}
    \subfigure[]{\includegraphics [scale=0.47]{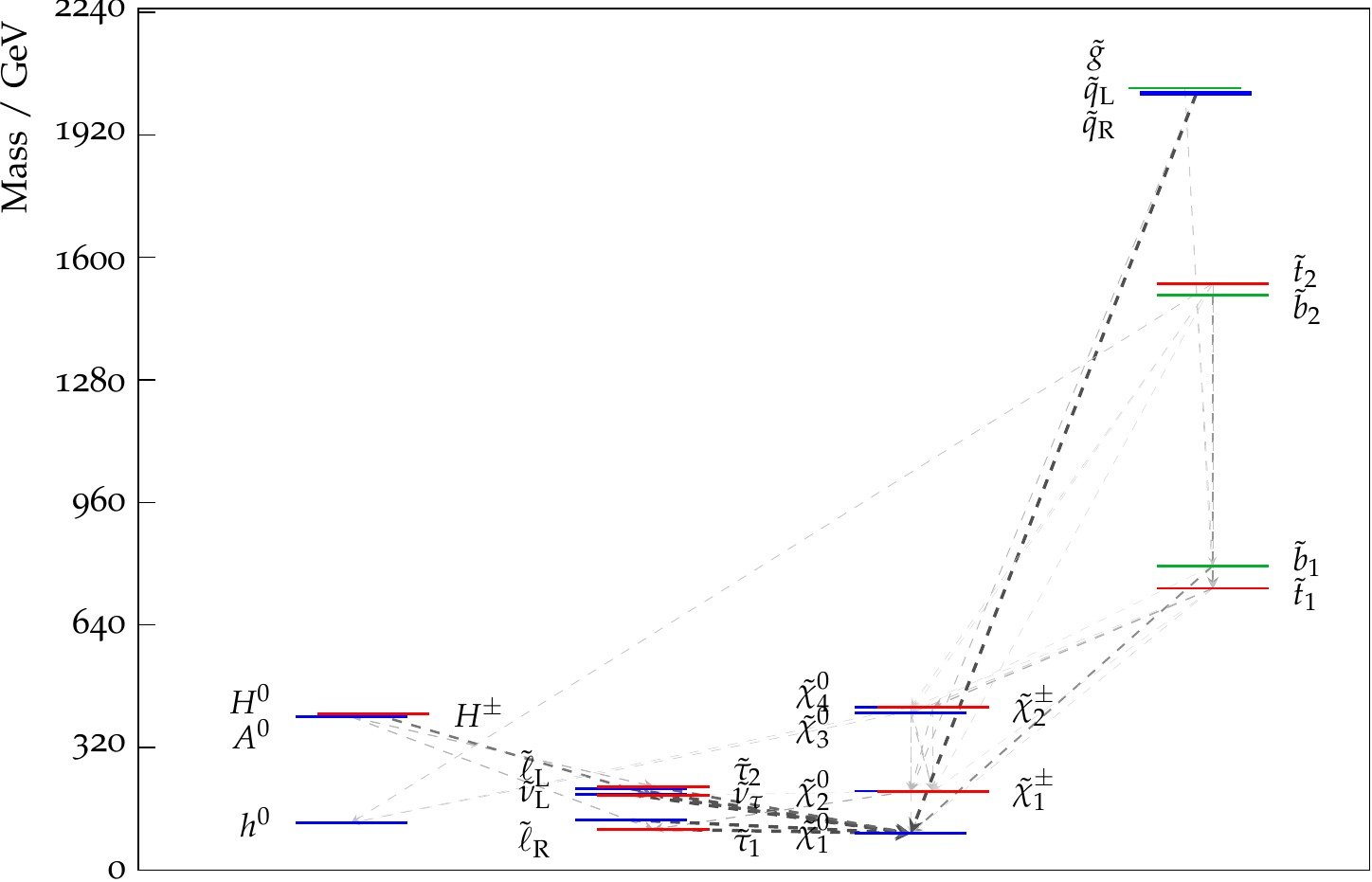}}
    \subfigure[]{\includegraphics[scale=0.25] {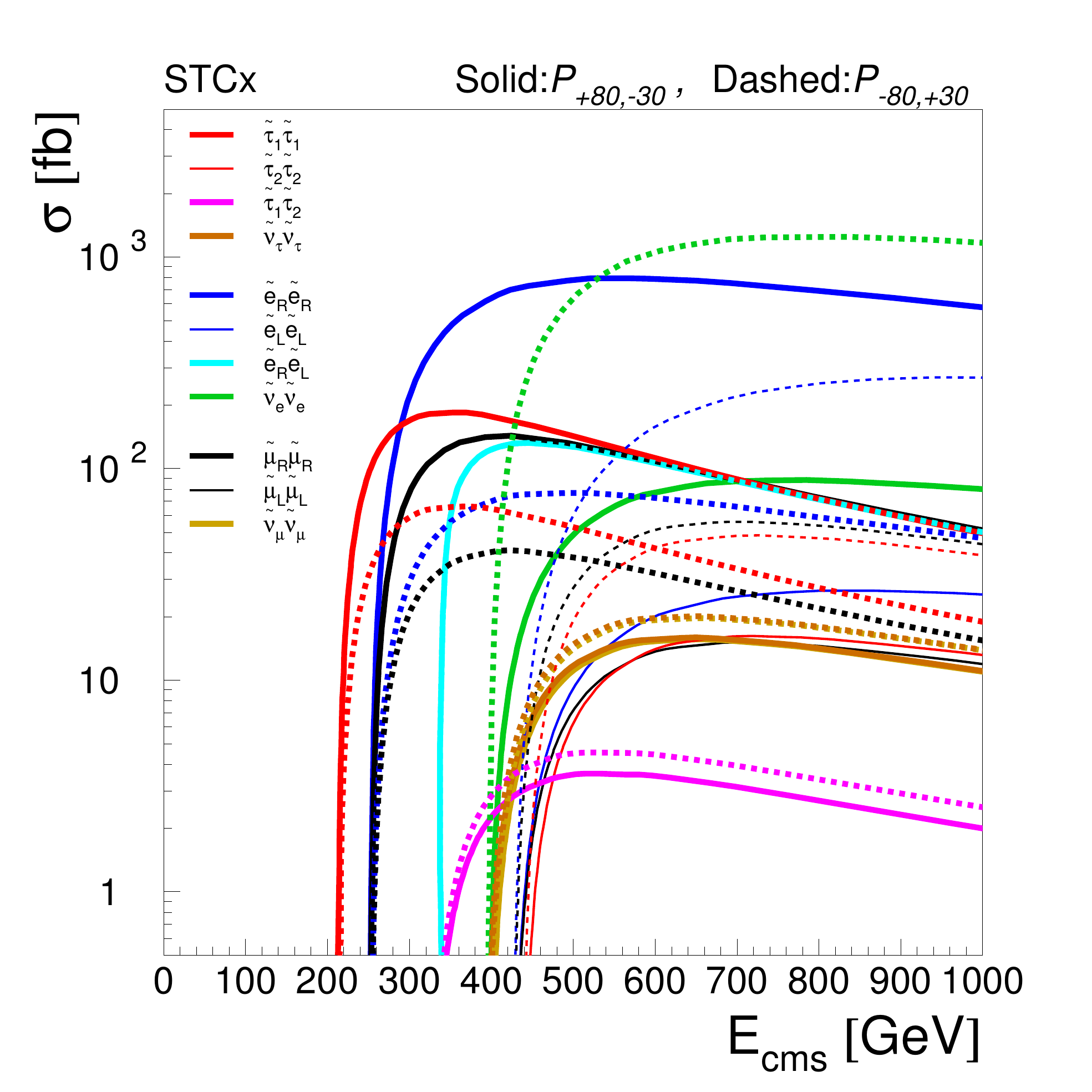}}
  \end{center}
  \caption{\label{fig:342_3} 
(a): Mass-spectrum and decay modes  (with BR $>$ 10 \%) of STC8. Heavier lines indicate larger
BR; 
(b): production cross-sections
   at ILC for sfermions, for $\mathcal{P}_{+80,-30}$ (solid) and $\mathcal{P}_{-80,+30}$ (dashed).
}
\end{figure}

A SUSY coannihilation model should -
in addition to containing a DM candidate -
accommodate the LHC limits,  the requirement from naturalness that the third generation squarks
should be comparatively light, and that the higgs-sector, in particular the higgs-mass, should
agree with observations.
It should also fulfil all constraints for precision measurements.
We found that this can all be achieved in a set of SUSY models with  12 parameters \cite{Baer:2013ula},
called STCx,
which we study in the following.
The different models (STC4 to STC8) differ
only in the magnitude of 
the $\stq$ mass,
while the bosino and slepton sectors are identical.
Figure \ref{fig:342_3}(a) shows the mass-spectrum of the STC8 model.
In \cite{Berggren:2015qua}, the models were  studied at LHC14 and ILC with the fast detector simulation programs.
The ILC study is the topic of the next sections, while a summary of
the findings at LHC is that 
the high mass first and second generation squarks and the gluino will not be seen even at the HL-LHC, since
  the cross-section is quite small.
However, the lower mass third generation squarks will eventually be discovered,
and  the bosino/slepton sector will be detected, but with little possibility to disentangle the different states.
\subsection{STC at ILC}

By inspecting figure  \ref{fig:342_3}, one sees that the expected signal
at a 500 \GeV~ ILC typically contains a few leptons and LSP:s,
leading to low multiplicity events. 
The events are central in the detector, and would have much missing energy.
Furthermore,   the cross-sections might be close to 1 pb for certain channels.
Heavier states often decays in cascades over  the NLSP, which is the $\stone$.
The mass difference between the $\stone$ and the LSP is  $\sim$ 10 \GeV, meaning that
the energy of the $\tau$:s from decays of  pair-produced $\stone$:s is  between 2.5 and  45 \GeV.
The SM background to such events is either processes with
real missing energy, e.g. $ZZ, WW \rightarrow \ell\ell\nu\nu$,
or with
``fake'' missing energy e.g. $\gamma\gamma$ processes, ISR, or single IVB, where
a high energy final-state particle escapes the detector through the low angle
acceptance holes due to the beam-pipes.

These considerations lead to the following selection criteria, valid
for all slepton studies: The events should contain less than  10 charged particles,
and two opposite charge lepton candidates (electrons, muons, or $\tau$-decay products). 
The total visible energy in the event, $E_{\mathrm{vis}}$ and the missing mass, $M_\mathrm{miss}$,
should be $<$  300 \GeV~ and $>$  200 \GeV, respectively, and all
particles should have momentum $<$ 180 \GeV.

An integrated 
luminosity of 1000 fb$^{-1}$ at $E_{cms}= 500$ \GeV,
evenly divided between
beam polarisations of
$\mathcal{P}_{+80,-30}$ and $\mathcal{P}_{-80,+30}$ was assumed.
Both signal and background were simulated with the
{\tt Whizard v1.95}~\cite{Kilian:2007gr} event-generator.
The response of the ILD detector ~\cite{Behnke:2013lya} was simulated with the fast
simulation program {\tt SGV}~\cite{Berggren:2012ar}.

To extract the masses of the SUSY particles, one uses that in the lab-frame, the highest and lowest
possible lepton energies are
\begin{align}
 E'_{\begin{smallmatrix}max\\ (min)\end{smallmatrix}} = 
\left ( \gamma \begin{smallmatrix}+\\ (-)\end{smallmatrix} \gamma \beta \right ) E_{rest~frame} = 
\frac{ E_{Beam}}{2}  \left ( 1 - \left( \frac{ \MXN{1}}{M_{\tilde{\ell}} } \right )^2 \right ) \left ( 1 
\begin{smallmatrix}+\\ (-)\end{smallmatrix}
\sqrt{1 -  \left ( \frac{ M_{\tilde{\ell}} }{ E_{Beam} } \right )^2} \right )
\end{align}
In this expression, there are two observables( $E'_{\begin{smallmatrix}max\\ min\end{smallmatrix}}$) and two parameters ($M_{\tilde{\ell}}$ and $\MXN{1}$).
For $\selr$ and $\smur$, $E'_{\begin{smallmatrix}max\\ min\end{smallmatrix}}$ can be measured very well at the ILC,
while only the end-point of the spectrum of  $E_{\tau-jet}$,
which is equal to  $E'_{max}$, can be well measured for $\stone$.
Therefore, we use $\selr$ and $\smur$ to determine $\MXN{1}$, and the end-point of the $E_{\tau-jet}$ spectrum to
determine $\mstone$.

\subsection{The selectron, the smuon and the LSP mass} 

As can be seen in Fig.~\ref{fig:342_3} (b),
the slepton pair
production cross sections are large in our scenario\footnote{
Due to the $t$-channel neutralino exchange, it is particularly large for the selectron channel.}.
Very precise measurements 
of the slepton and LSP masses can be obtained from the edges of the lepton spectra.
In addition to the general slepton selection criteria,
it is demanded that the two lepton candidates have the same flavour.
To separate the right-handed and left-handed states, switching of the beam-polarisation is helpful,
as it will preferentially select the process  either with left-handed or right-handed slepton production.
The left-handed states are heavier than than right-handed ones, meaning that the different kinematics
can be used to 
further enhance the separation. 
Finally, a 
``Tag and probe'' technique is used: a lepton-candidate is accepted only if {\it the other} candidate
in the event has a momentum within the kinematic limits of the process.
With these cuts, the selection efficiency is 51\,\% for $\selr$,
while it is slightly higher,  65\,\% 
for $\smur$.


\begin{figure}[htb]
  \begin{center}
    \subfigure[]{\includegraphics[scale=0.20] {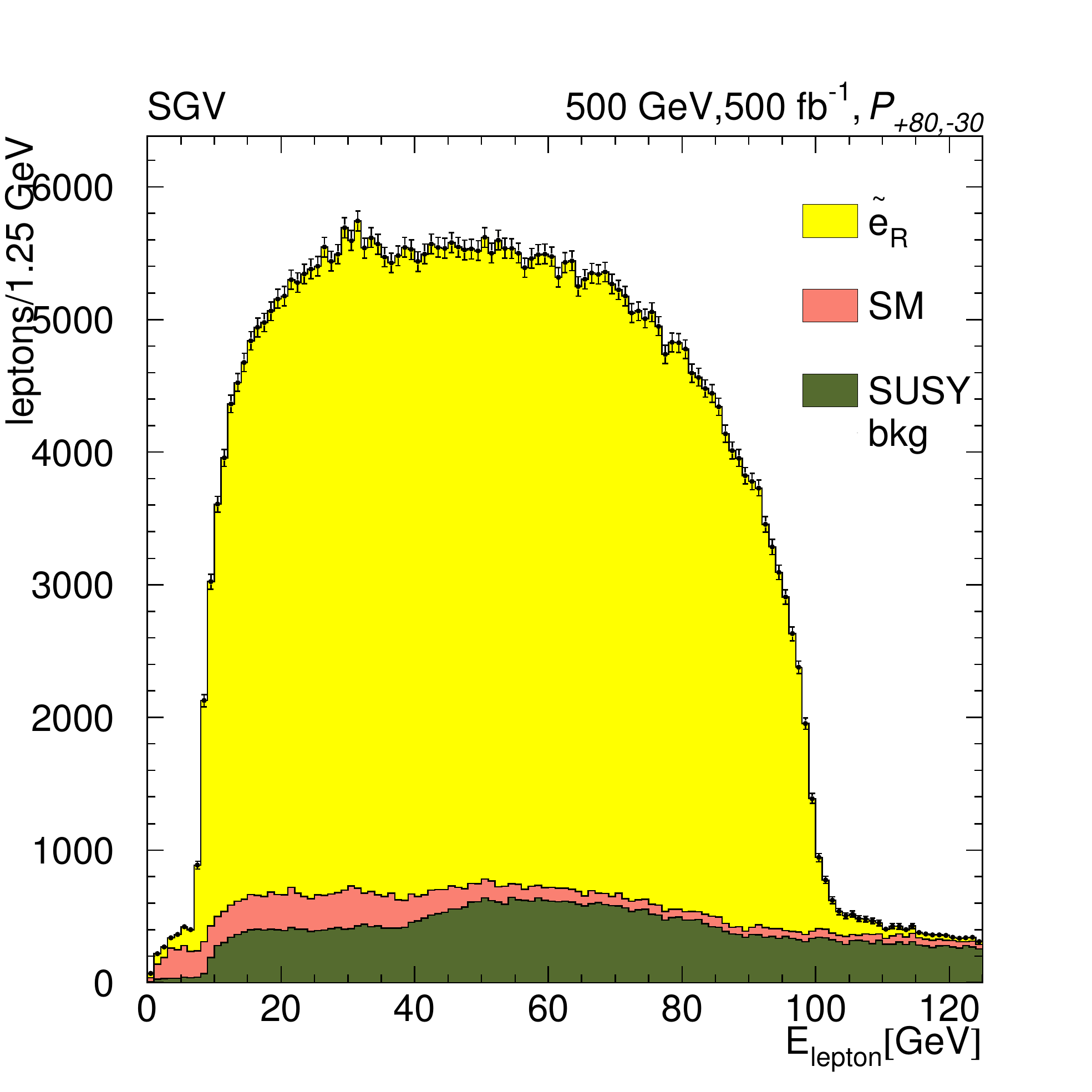}}
    \subfigure[]{\includegraphics[scale=0.20] {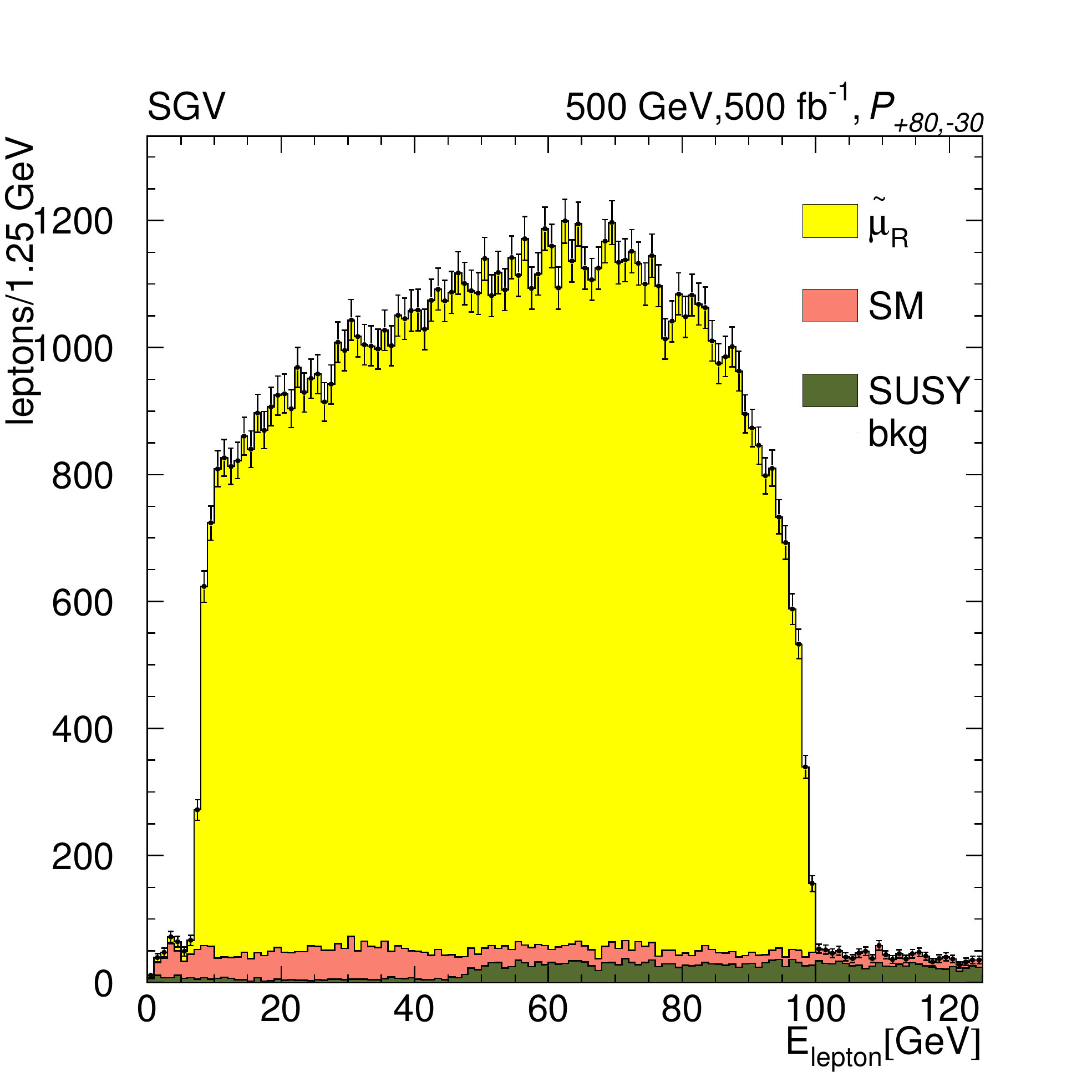}}
    \subfigure[]{\includegraphics [scale=0.20]{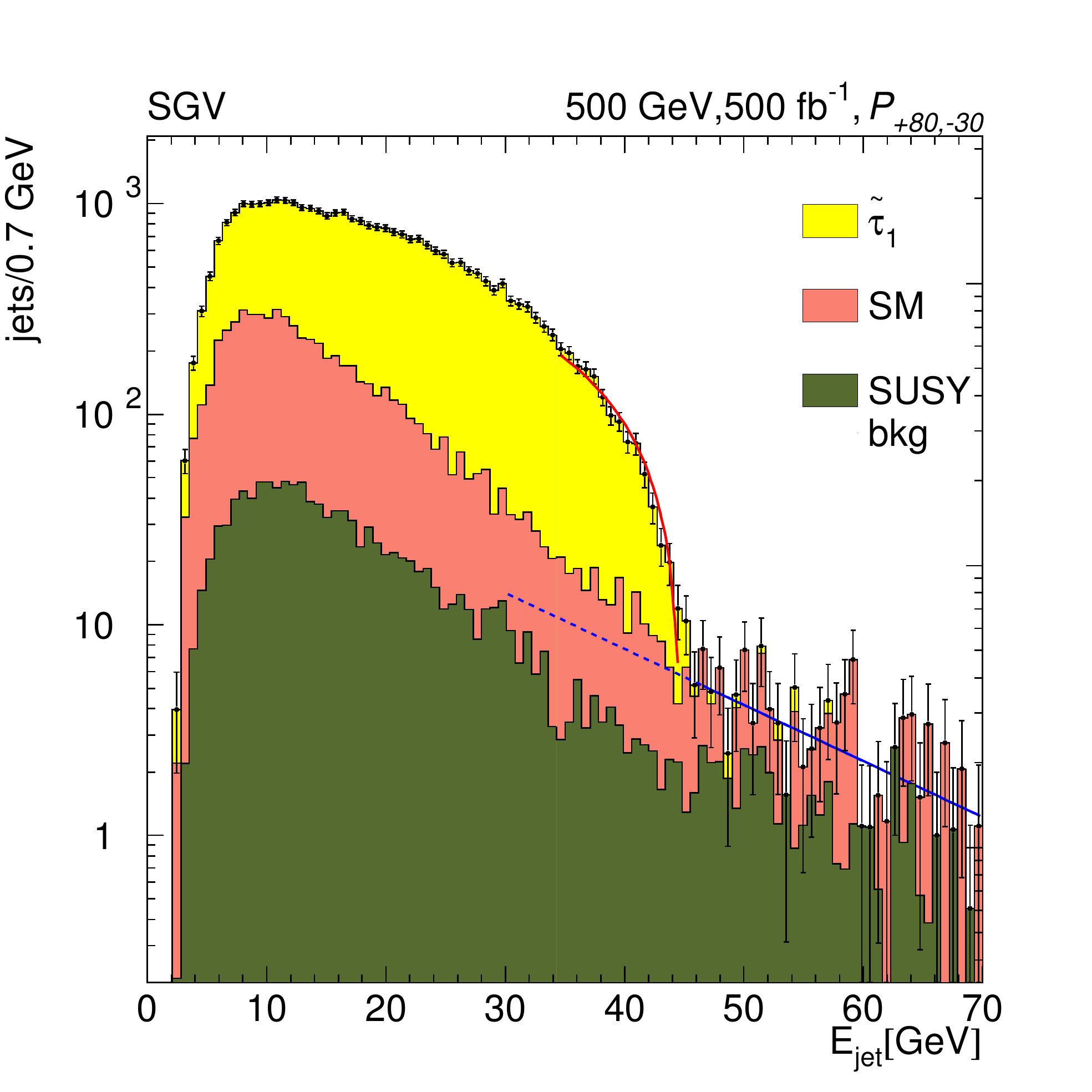}}
  \end{center}
  \caption{\label{fig:342_5} Spectra of electron (a), muon (b) and $\tau$-jet (c) energies in selected di-leptons events
after collecting
500 fb$^{-1}$ of data for beam-polarisation  $\mathcal{P}_{+80,-30}$. }
\end{figure}

Figure~\ref{fig:342_5} (a) and (b)
shows the spectra in selected di-leptons events.
The signal stands out above quite small SM and SUSY background.
By fitting the two edges in each spectrum, and using eq. (3.1), we obtain
$\MXN{1}   = 95.47  \pm 0.16 ~\GeV$ and $ M_{\selr} = 126.20 \pm 0.21 ~\GeV $ from the $\selr$ spectrum,
and $\MXN{1} = 95.47 \pm 0.38 ~\GeV$
and $M_{\smur} = 126.10 \pm 0.51 ~\GeV$ from the $\smur$ spectrum.
This is in good agreement with the model-values:
the true masses in STCx are  $\MXN{1} =95.59 ~\GeV$, $M_{\selr} = 126.24 ~\GeV$ and  $M_{\smur} = 126.16 ~\GeV$.
Combining the measurement of the LSP mass from these two  analyses
gives an uncertainty of $\sigma_{\MXN{1}} = 147 ~\MeV$,
i.e. slightly above 1 per mil.

\subsection{The stau mass and the mixings} 
%
As was pointed out above, a precise determination of the 
$\stau$ sector is essential to test whether the $\XN{1}$ is the main constituent of
DM.  
Since the $\stau$ mass difference to the LSP is much smaller, and the $\tau$ decays before detection,
the spectrum is softer than that of the leptons
from \selr~ or \smur~ decays. 
In addition, also due to the decay of the $\tau$:s, particle identification becomes less effective in
suppressing the background.
The signal therefore resembles $\gamma\gamma$ to a much larger extent,
and also resembles $WW$ or $ZZ$ events decaying to $\tau\nu\tau\nu$.
Several further criteria must therefore be added:
The requirements on $E_{\mathrm{vis}}$ and $M_\mathrm{miss}$ are strengthened
to $<120 ~\GeV$ and $> 250 ~\GeV$, respectively, and
the visible mass
should be 5 \GeV~ below $M_{Z}$,
To further reduce the  di-boson background, as well as the SUSY background from  \selr- or \smur-pair production,
it was demanded that the candidate events are not same-flavour lepton ones.
To reduce the $\gamma\gamma$ background,
the direction of the missing momentum is required to be more
central ($|\cos{\theta_{miss}}| < 0.8$),
$M_\mathrm{vis}$ to be above 20 \GeV, and
the total energy observed below
30 degrees to the beam-axis to be below 2 \GeV.
A cut on the likelihood that the event is a $\gamma\gamma$ event is also imposed.
With these cuts, the selection efficiency
for $\stone$-pair production is 17\,\%.

The end-point of the spectrum of the $\tau$ decay-products
is determined by fitting the background in the signal-free region
well above the endpoint,  and then fitting the excess above the extrapolated background fit
to a signal contribution.
Figure~\ref{fig:342_5} (c) shows the energy-spectrum 
of selected $\tau$-jets.
The endpoint
could be determined to be E$_\mathrm{endpoint}=44.49^{+0.11}_{-0.09}$ \GeV,
corresponding to an uncertainty on $M_{\stone}$ of 200 \MeV, if an uncertainty on
the LSP mass from the $\sel$ and $\smu$ analysis of $\sim$ 100 \MeV~ is assumed.
Hence, also $M_{\stone}$ can be determined to slightly above 1 per mil.

In~\cite{Bechtle:2009em}, where a model quite similar to STCx has been studied,
it is found that, in addition to $M_{\stone}$, the cross section can be
determined at the level of 4\,\%, and the polarisation
of $\tau$-leptons from the \stone~ decay, which gives access to the $\stau$ and 
$\XN{1}$ mixing,
could be
measured with an accuracy of 5\,\%.
For the current study, the corresponding analysis is work-in-progress, but it seems quite probable that
also mixings would be measurable at the required percent level.
\section{Conclusions}
SUSY models with a rich and compressed spectrum
are still the best fit to data. They are not excluded by LHC.
It is likely that LHC would discover such a model in the next few years, if it is realised in nature.
In such models, a rich spectrum is reachable by the ILC,
  and ILC will be able to corroborate on any discovery made at LHC.
In particular, ILC will be able to prove that the New Physics
  discovered at LHC is SUSY. Masses will be determined at per mil-level,
  mixings (probably) at percent-level.
With such precisions, ILC will be capable to measure DM properties with an accuracy
  close to the cosmological observations on  CMB from the Planck satellite,
and hence to determine if indeed the LSP is the main contributor to DM.

\bibliography{
berggren-ichep-chicago-aug-2016-proceedings}
%
%
\end{thebibliography}

\end{document}